\documentclass[conference]{IEEEtran}
\IEEEoverridecommandlockouts
\usepackage{cite}
\usepackage{amsmath,amssymb,amsfonts}
\usepackage{algorithmic}
\usepackage{graphicx}
\usepackage{textcomp}
\usepackage{xcolor}
\def\BibTeX{{\rm B\kern-.05em{\sc i\kern-.025em b}\kern-.08em
    T\kern-.1667em\lower.7ex\hbox{E}\kern-.125emX}}

\begin{document}

\title{Reactive Failure Mitigation through Seamless Migration in Telecom Infrastructure Networks}

\author{
	\IEEEauthorblockN{Mahdi Soleimani and Vahid Shah-Mansouri \\
		\IEEEauthorblockA{School of ECE, College of Engineering, University of Tehran, Iran \\
			Emails:\protect  \{mahdi.soleimani,  vmansouri\}@ut.ac.ir}}
}

\maketitle
\vspace{-10mm}
\begin{abstract}
Various methods are proposed in the literature to mitigate the failures in an infrastructure network. Failure mitigation can be carried out in an active or passive manner. In active manner, live backups provide the required reliability. Due to the high cost of active backups, failures can be mitigated in a reactive manner where mitigation starts when a function fails. In this paper, network functions are divided into two classes of essential (i.e., core) functions and additive (i.e., service) functions. Core functions need active backups and cannot tolerate any failure. However, additive functions do not require backups and their absence is tolerable in short periods. Maximum tolerable (function) absence time (TAT) is a measure used for their reliability level. In our model and within the mitigation process, first, the backup host to migrate the failed function is selected. Then, the state of the function is migrated. The process should be carried out  in less than TAT. An optimization problem is formulated investigating cost-effective failure mitigation of additive functions.  The problem is of mixed integer non convex form. To tackle the computational complexity, it is bisected into two back-to-back leader-follower parts. The leader selects the migration destinations and the follower divides network resources among functions for migration. These problems are both NP-hard. For the leader problem, a central heuristic based on the Viterbi algorithm suggested. The follower problem is converted into two convex sub-problems. The validity of the proposed algorithms is proved via extensive numerical results.
\end{abstract}


\section{Introduction }
An infrastructure provider (IP) continuously accepts service requests as forwarding graphs with required functions, their respective required resources, essential network connections among these functions, and their reliability requirement. 

The predominant scheme for guaranteeing customer desirable reliability identifies a reliability measure as an uncorrupted work probability $(1 - Pr\{\text{Failure}\})$
for each function \cite{Fan:2015:GGR:2785989.2786000}. One solution to meet the reliability requirement is to deploy active and live backups where the state of the backup functions are synched with the main function. Keeping simultaneous state-aware active backups in the chain requires IPs to assign redundant resources and keep them active in a significant portion of time to attain lower failure probability. For example, \cite{7996840} tries to optimize this redundant cost by changing in service instance of a function and service routing rules in the case of a rare host/function failure occurrence. In \cite{7969152}, authors tried to minimize the number of backup nodes.

Modern network infrastructure providers are frequently involved in resource pool insufficiency. Indeed, tremendous expenses of resource pool update and power consumption expenses in data centers may lead to a high service request rejection rate. Therefore, the IP becomes gradually disable of assigning enough backup resources to every network function. To overcome ineffective resource preoccupation for backing up every single existing function, exposed functions in an FG are divided into two classes, essential (core) functions, and additive functions. Failure of the core functions is not tolerable by the IP customer, even in the short periods, whereas additive function failures considered acceptable in small time spans. An IP customer (i.e., a service provider) specifies types of functions and their respective reliability measures in the SLA. For a core function, this measure consists of impeccable acting probability. In contrast, for an additive function this information rather identifies maximum tolerable access time (TAT). TAT shows the maximum failure recovery time the customer can tolerate.

Process state information is the set of program memory components which are required for the process execution integrity and consistency. For an additive function to recover seamlessly, IP should first appoint a node in the infrastructure for the function to migrate there. Then, the process state information should be exchanged in a period that lasted less than TAT to the specified destination. In \cite{7274710}, authors proposed a method to decrease machine state diffusion time in the network, however the time of state information reception at the destination is not surveyed. In \cite{6546155}, virtual network functions (VNFs) arrangement in the network is changing dynamically to optimize the network power consumption, after network state transition, e.g., by service departures. However, the cost of imposing redundant migrations to the network and migration power consumption are neglected. \cite{7883907} aimed to manage power consumption too, but unlike \cite{6546155} noted the trade-off between migration consumed power and  services power usage. In \cite{7936622}, authors tried to minimize migration time in the network by determining the migration order of the VNFs and transmission rates in the network. In \cite{7500390}, vCDN migration time is optimized.

In this paper, we propose a failure mitigation process. The failure mitigation process includes two parts. A destination host is selected for the the migration of the failed function. Also, the state information is migrated from the failed host to the selected destination. We formulate this problem as a optimization problem. The objective of the problem would be to minimize the network and forwarding costs that would be imposed on the IP, where the migration time is kept less than or equal to the TAT measure for each VNF. The algorithm specifies the VNFs respective migration destinations as well as how the IP should use the idle network resources to migrate the ascertained state information for each VNF. Thus and so, the IP would be able to handle network failures efficiently. 

The rest of the paper is organized as follows. In Section ...


\section{System Model}
Consider a directed graph expressed by $G(E, N)$ as the infrastructure network, where $E$ is the set of links and $N$ is the set of nodes of the IN. Each node in $N$ has some idle computing resources equals to $R_c(n)$ and link $e$ in the network has a nominal transmission rate, $B(e)$, and $B^v(e)$ is the idle resource of link $e$. 
Let $P$ denote the set containing paths of the graph. This set can be obtained in polynomial time using the set $E$. $F$ is the set of the failed VNFs including $|F|$ VNFs. The parameters for each member of $F$ is represented in Table \ref{vnfspecs}.

\begin{table}[t]
	\caption{Failed VNFs Specifying parameters}
	\begin{center}
		\begin{tabular}{|c|c|c|}
			\hline
			\textbf{Parameter}&\multicolumn{2}{|c|}{\textbf{Parameter Specification}} \\
			\cline{2-3}
			\textbf{variable} & \textbf{\textit{Range}}& \textbf{\textit{Declaration}}\\
			\hline
			$V(f)$& $\mathbb{R}^+$ & VNF state information size  \\
			\hline
			$T(f)$& $\mathbb{R}^+$ & VNF TAT  \\
			\hline
			$C(f)$& $\mathbb{R}^+$ & VNF required processing resources \\
			\hline
			$R(f)$& $\mathbb{R}^+$ & VNF input data rate \\
			\hline
			$S(f)$& $\in N$ & Failure point of the VNF \\
			\hline
			$L^+(f)$& $\in N$ & Next N-PoP in VNF's SC \\
			\hline
			$R^+(f)$& $\mathbb{R}^+$ & VNF's outgoing rate \\
			\hline
			$L^-(f)$& $\in N$ & Previous N-PoP in VNF's SC \\
			\hline
		\end{tabular}
		\label{vnfspecs}
	\end{center}
\end{table}
The migration destination of each VNF, $D(f)$, is the output of the problem where $\mathbf{D}$ is the vector of destination hosts. A $|F| \times |E|$ matrix is defined named $B$ to model the procedure of using idle network resources for migration. Each element of this matrix, $\beta_{fe}$, is a positive number less than one and shows the proportion of idle resources on link $e$ which is used by function $f$ for state migration. IP disseminates failed function state load among multiple paths. Let $\alpha_{fe}$ denote the proportion of the failed function $f$ state information transmitted through link $e$ during the migration process. Thus, matrix $A$ is defined which its $fe$ element is $\alpha_{fe}$. The selected destination should substitute the primitive host and reconnect the service chain. Therefore, the program should select directed paths connecting $L^-(f)$ to $D(f)$ and $D(f)$ to $L^+(f)$ to do so. Provided that the outgoing and incoming rate constraints of the chain could be added to settled traffic of the network without violating bandwidth limits. These paths, denoted by $\rho^-(f)$, and $\rho^+(f)$, respectively, are stored in $\mathbf{P}^+$, and $\mathbf{P}^-$ matrices.

Time for migration of a VM is accumulation of four parts as
\begin{align}
t_{\text{migration}} = t_{\text{algorithm}} + t_{\text{startup}} + t_{\text{dissemination}} + t_{\text{retransfer}}. \label{eq1}
\end{align}
The first part, $t_{\text{algorithm}}$, enfolds the time which algorithm needs to decide on the output parameters before starting the migration phase. The problem is NP-hard, so dealing with suboptimal fast-executed heuristics seems inevitable. We neglect this time for the heuristic algorithm. The second part, $t_{\text{startup}}$, encompasses the time that the system devotes to prepare hardware hosting resources in the destination, considering the startup time. The third and perhaps the central portion, $t_{\text{dissemination}}$, includes the state dissemination time. With the assumption of parallel execution of load dissemination and resource preparation and the plausible presumption that dissemination time is much more considerable, only dissemination time would be substantial. Uninterrupted service and seamless failure mitigation require IP to capture the packets of the failed function during the mitigation phase and forward them to the selected destination at the end. Thus, the mitigation is not finalized just after the dissemination phase’s is finished. The final part of the delay, $t_{\text{retransfer}}$, is required in order to retransfer these packets. Therefore, the approximate aggregate migration time would be as expressed in equation \eqref{eq1}. With the reasonable assumption that the resources of the network for retransfer and state dissemination is proportional, then retransfer time would be modeled inside dissemination time by changing function input parameters, including TAT, $T(f)$, and state information size, $V(f)$. In overall, an approximate migration time would be as
\begin{align}
\label{eq2}
t_{\text{migration}} \approx t_{\text{dissemination}}.
\end{align}

Dissemination time is 
maximum single path traverse time. For the first packet which is transferred, end-to-end time is equal to the aggregation of the transmission delay, the queuing delay, the propagation delay and the processing delay 
along the end-to-end path.  There is no specific processing involved in the migration process.
Indeed, there are usually as many network resources accessible in the datacenter IP networks as the propagation delay becomes negligible too, and the link transmission rate is as high as the transmission delay would be insignificant too. The last state information packet of agent $f$ in link $e$ sees maximum queuing delay as $V(f)\alpha_{fe}/(B^v(e)\beta_{fe})$, so an agent dissemination time is the maximum load discharge time on the links which 
function $f$’s information have passed as
\begin{align}
t_{\text{migration}}(f) \approx t_{\text{dissemination}}(f) \approx \max_{e \in E}^{}(\dfrac{V(f)\alpha_{fe}}{B^v(e)\beta_{fe}})\label{eq3}
\end{align}

\section{Cost Function}

For the IP, 
 reserving network and processing resources is an internal action and cause no excessive charge. According to \cite{7823768}, the main cost endured by the IP to migrate VMs in a network is the cost of holding the packets in the central controller and forwarding of these packets towards the recovered function. 
  Moreover, to abstain the overuse of network resources in order to avoid long queues, another cost components is used as
\begin{align}
C_{\text{total}} &=  \omega_{\text{forwarding}}\times C_{\text{forwarding}} + \omega_{\text{congestion}}\times C_{\text{congestion}} \label{eq6}
\end{align}
\noindent where $\omega_{\text{forwarding}} + \omega_{\text{congestion}} = 1$ are design parameters.

\subsection{Controller Forwarding Cost}
IP would be charged linearly for every agent forwarded traffic to the central controller. The total traffic forwarding cost would be as follows. $c_u(f)$ is unit forwarding cost of the agent $f$, $R(f)$ equals to the forwarding rate of this agent, and $t_{\text{migration}}(f)$ pasted as shown in \eqref{eq3}. Then, we have
\begin{align}
C_{\text{forwarding}} &=  \sum_{f \in F}(c_u(f)\times\max_{e \in E}\left\{\dfrac{V(f)\alpha_{fe}}{B^v(e)\beta_{fe}} \right\} \times R(f)) \label{eq4}
\end{align}
\subsection{Congestion Cost}

To model the congestion cost of a single link, the queueing function $\dfrac{\mu}{\mu - \lambda}$ is used where $\mu$ is the nominal link rate, and $\lambda$ is the link servicing rate \cite{Bertsekas:1992:DN:121104}. Then, we have
\begin{align}
C_{\text{congestion}} &= \sum_{e \in E}c_b(e)\times\dfrac{B(e)}{B^v(e).(1 - \sum_{\phi \in F}^{}(\beta_{\phi e}))} \label{eq5}
\end{align}
In \eqref{eq5}, $c_b(e)$ is the link unit-lengthed queue cost, and the fraction on the right side shows queue length. The link nominal capacity is $B(e)$ and its remaining capacity during migration is $B^v(e).(1 - \sum_{\phi \in F}^{}(\beta_{\phi e}))$.

\section{Problem Constraints}
In this section, the constraints of the problem is presented.
\subsection{Flow Conservation Constraints}
Failed VNFs traffic should conserve across network nodes so elements of $\mathbb{A}$ matrix should satisfy
\begin{align}
\quad \sum_{\forall e \in E}(\alpha_{fe} (D_b(e, n) - S_b(e,n)))=&\left\{\begin{array}{rcl} 1 & ; & n = D(k), \\ -1 & ; & n = S(k), \\ 0 & ; & o.w. \end{array} \right. \label{eq7}
\end{align}
$D_b(e, n)$ is a binary variable which returns one if the node $n$ is the incoming end of the link $e$, $S_b(e, n)$ is doing so for outgoing nodes.
Equation \eqref{eq7} expresses that the difference between incoming and outgoing $\alpha$ flux in a node should be -1, 1 or 0, for an agent source host, $S(f)$, destination host, $D(f)$, or middle traversing point, respectively.

Elements of $\mathbb{B}$ matrix should satisfy below constraint.
\begin{align}
\quad \sum_{\forall e \in E}(\beta_{fe} (D_b(e, n) - S_b(e,n)))&\left\{\begin{array}{rcl} > 0 & ; & n = D(F_k), \\ < 0 & ; & n = S(F_k), \\ = 0 & ; & o.w. \end{array} \right. \label{eq8}
\end{align}
Equation \eqref{eq8} expresses that difference between incoming and outgoing links idle capacities devoted to agent $f$'s flow should be positive, negative or zero for an agent source host, $S(f)$, destination host, $D(f)$, or middle traversing point, respectively.

\subsection{Service Chains Establishment Constraints}

Selected paths for chain establishment after migration, $\rho^+(f)$, and $\rho^-(f)$ should have enough capacity to provide $R^+(f)$ and $R(f)$ for every traversing chain. This This can be written as a constraint as
\begin{align}
\ \sum_{f \in F}^{}[\Omega(e, \rho^+_{f}).R^+(f) + \Omega(e, \rho^-_{f})R(f)] \leq B^v(e), \ \forall e \in E\label{eq9}
\end{align}
\noindent where $\Omega(e, p)$ is a binary variable 
 equals to one if path $p$ traverses link $e$. The constraint provides that added load on a link after migration will not exceed network resources limit on that link, i.e., $B^v(e)$.

\subsection{Resources Limitation Constraints}

Adequate processing resources should be present in end-host nodes, $D(f)$ as 
\begin{align}
\sum_{f \in F}^{}(\delta(n - D(f))\times C_r(f)) \leq R_c(n), \ \forall n \in N. \label{eq10}
\end{align}
Delta function shows whether or not the node is selected destination by agent $f$ . Requested network resources for migration, which is the summation of idle bandwidth proportions used by set $f$ members, should not exceed available bandwidth on the respective link, i.e., $B^v(e)$ as
\begin{align}
	\forall e \in E: \sum_{f \in F}^{}\beta_{fe} \leq 1 \label{eq11}
\end{align}
\subsection{Migration Time Constraints}
As explained in previous section, with respect to each agent's TAT, the following constraint should be satisfied.
\begin{align}
t_{\text{migration}}(f) \approx \max_{e \in E}^{}\left\{\dfrac{V(f)\alpha_{fe}}{B^v(e)\beta_{fe}}\right\} \leq T(f), \ \forall f \in F. \label{eq12}
\end{align}

\subsection{Complexity Analysis}
Members of vector $\mathbf{D}$ are integers
so the problem is of mixed-integer form. All constraints, except \eqref{eq12}, are linear. This constraint can omitted without much loss of generality. It is because firstly the presence of the forwarding part in the cost function results from IP to decrease the weighted sum of migration times as much as possible. Secondly, in reality, machines migrate before failure occurrence and could attune their migration trigger time so that the control process shall have enough time for after shut-down seamless migration. Congestion cost is directly obtained from a convex function 
and therefore it is convex. However, the forwarding part of the cost is not convex.
Therefore, the whole problem turns into the mixed integer non-convex programming form and is NP-hard to solve. To conquer complexity, the problem is divided into two sequential leader and follower sub-problems, and it is solved.

\section{Leader Problem}\label{leader}
The leader problem deals with the destination selection part of the problem. Input parameters are those mains' input parameters which affect destination selection, including migrating VNFs set, $F$, and their determining parameters, $S(f), R(f), L^-(f), R^+(f), L^+(f), C_r(f)$, as well as infrastructure network structure and conditions. i.e. $G(E, N), B(e), B^v(e), R_c(n)$, plus whether or not a specific server is idle, which is denoted by $\iota(n)$ binary variable.
\subsection{Leader Problem Utility Function}
The leader problem utility function is composed of three parts as explained in \eqref{eq0}. $\omega_n + \omega_l + \omega_p = 1$, $U_n, U_l, U_p$ defined in \eqref{eq14}, \eqref{eq15}, \eqref{eq16}.
\begin{align}
	max(\omega_n.U_n + \omega_l.U_l + \omega_p.U_p) \label{eq0}
\end{align}
\subsubsection{Migration Resources Utility}
The main goal of the leader problem is not to confine network resources, which are given to follower problems for migration. So that the feasible region of the follower problem, according to the TAT constraints 
does not shrink.
Therefore, the first part of the utility function 
aimed to maximize multi-path E2E network resources between source and destination nodes. $\chi(n, m)$ is a function that shows E2E aggregated multi-path capacity between nodes n and m. This function is determinable using adjacency matrix of the network and its limited order powers in polynomial time.
\begin{align}
U_n = \left\{ \dfrac{  \sum_{f \in F}(\chi(S(f), D(f))   }{\sum_{f \in F}(\max_{\forall n \in N}(\chi(s(f), n)))} \right\}. \label{eq14}
\end{align}
The numerator of the fraction in \eqref{eq14} shows aggregated E2E capacity between source and selected destination hosts, while the denominator expresses aggregate E2E bandwidth between sources and best choices among destination candidates — the fraction used to normalize the utility, so it would not cancel other utility parts.

\subsubsection{Load Balancing Utility}
The IP tries to use more of the servers with higher available processing capacities rather than other preoccupied servers. This part of the utility function is expressed as \eqref{eq15}, where $\vartheta(n)$ is a binary variable that shows whether or not node $n$ used as a migration destination. The utility function is
\begin{align}
 U_l = \dfrac{\sum_{n \in N}(\vartheta(n)\times R_c(n))}{\sum_{n \in N}^{}R_c(n)}. \label{eq15}
\end{align}
\subsubsection{Power Saving Utility}
To manage power consumption, IP tries to avoid using idle servers as much as possible by maximizing the relative number of not used idle servers, as described in \eqref{eq16}. The numerator counts the number of unused idle servers, by multiplying being idle indicator, $\iota(n)$ into not being used factor, $1-\vartheta(n)$. $N_{\text{idle}}$ shows number of idle servers in the network, i.e. $\sum_{n \in N}^{}\iota(n)$. This utility can be written as
\begin{align}
	U_p =\dfrac{\sum_{n \in N}([1 -\vartheta(n)]\times \iota(n))}{N_{\text{idle}}} \label{eq16}
\end{align}
Even without considering utility functions, the leader problem in its simplest form will be reduced to a bin packing problem, which is a combinatorial NP-hard problem.
\subsection{Viterbi Based Heuristic}
Viterbi is an algorithm used to find the most probable set of hidden events after observing an outcome sequence. This algorithm widely used in industry to decode convolutional codes. 
The algorithm needs to defined state and observation spaces, transition and emission matrices, and initial state probability vector to work \cite{1450960}.
\subsubsection{Crafted State and Observation Spaces}
Consider the set of network nodes as the state space and the set of network VNFs as the observation space. The failed VNFs encompasses the observation sequence. A trivial migration priority indicator suggested as follows in \eqref{eq17}. 
Evidently, a VNF priority should be directly dependent on its state information size, and rate and inversely dependent on it's TAT. As much as the TAT measure is high, it means the network has more time to mitigate its failure.
\begin{align}
\dfrac{R(f) \times V(f)}{T(f)} \label{eq17}
\end{align}
Obviously, the most important VNF, according to this parameter, is the first one in the failed VNFs tour. Then, IP should choose the VNF, which is most affected by the previous VNF selection among failed VNF set. Perhaps, the nearest VNF, respective to S(f) measure, to the previous one would be so and should be placed next to the previous one in the tour. Then, IP traverse along the failed VNF's set continuously by going to the nearest neighbors and would stop as soon as all the VNFs are visited, and the observation sequence is entirely defined. 
The algorithm should specify another turn as the set of selected destinations as hidden chains of events in the state space.
\subsection{Emission \& Transition Matrices}
Each element of the emission matrix, $E_{ij}$, shows the probability of observing $o_j$ from state $s_i$, i.e., the probability of a specific VNF being failed conditioned to a particular node being selected as a destination node. To determine this probability, some utility functions used to model the probability of a specific point selection as host by a specific failed VNF and then Bayes' theorem used to calculate desirable pseudo-probability. The utility function in \eqref{eq18} reflects three measures in the leader problem utility function. In this function, $\omega_b$ is the weight of the network resources utility, $\omega_c$ defines the weight of the load balancing utility, and $\omega_i$ used to utilize non-idle host more probably. $C_{set}(f) $ is a set of network nodes that satisfy chain establishment and resource limit constraints for function $f$. 
\begin{align}
\forall n \in C_{set}^f : U(n, f) = \omega_b.(\dfrac{\chi(S(f), n)}{{V(f)}/{T(f)}}) &+
\omega_c.(\dfrac{R_c(n)}{C_r(f)}) \notag \\ &+ \omega_i.(1- \vartheta(n)) \label{eq18}
\end{align}
In order to craft plausible probabilities, these utilities would be normalized to their summation over candidates set, $C_{\text{set}}(f)$.
Then desirable conditional probability would be calculated using Bayes' Theorem, and the fact that $\text{Pr}{\text{f being failed}}$ is fixed in all probability expressions and can be omitted, which requires final probabilities to be normalized once again at the end. Finally, the probability of a specific node to be selected as destination host in general unconditioned terms should be calculated, which means the occurrence chance of an internal state, without any external observation except the knowledge of existing one and only one hidden observation should be calculated. Given that the failure probability of distinct VNFs is equal, this probability would be expressed as follows in \eqref{eq20}. $\Upsilon$ is the set of all the VNFs in the network. $\digamma$ is the random variable points to the failed VNF, uniformly distributed over $\Upsilon$ set with mass probabilities $\dfrac{1}{|\Upsilon|}$.
\begin{align}
\text{Pr}\{D(\digamma) = n\} = \sum_{u \in \Upsilon}^{}(\dfrac{1}{|\Upsilon|}\times \text{Pr}\{D(u) = n \mid \digamma = u\} ) \label{eq20}
\end{align}
The transition matrix would be obtained with the same techniques used for emission matrix calculation. But with different utility parameters reflects the rational behavior of adjacent VNFs in the observation sequence, respecting the resource limit, and step-by-step updated chain establishment constraints. (Used resources by more important functions obtained from the resource pool after each step.)

\section{Follower Problem}\label{follower}
\subsection{Objective Function Further Simplification}
The follower problem should specify $\mathbf{A}$ and $\mathbf{B}$ matrices. The cost function of the problem is the same as the primary problem cost function. Wherein, the forwarding part would conquer the congestion part, if the latter part is kept in the almost-linear part of $\dfrac{1}{1-x}$ function. However, when the congestion part goes to the asymptotic infinite region, $\dfrac{\lambda}{\mu}> 0.9$, the congestion part becomes the determining factor. This would cause the whole problem to behave inconsequently, paying no attention to TAT limits, and threatening central controller proper functioning by overloading it to avoid asymptotic costs of congestion. To avoid such an intricate situation, the congestion part of the cost function should be omitted, with controlling the congestion cost by updating network limit constraints as
\begin{align}
B^{\prime}_{v}(e) = \max\left\{ B^v(e) - \dfrac{B(e)}{10}, 0 \right\} \label{eq21}
\end{align}
Then, the follower problem cost function reshapes as
\begin{align}
C_{follower} &= \sum_{f \in F}^{}[c_u(f) \times R^{\prime}(f). \max_{\forall e \in E}(\dfrac{V(f).\alpha_{fe}}{B^v(e). \beta_{fe}})] \label{eq22}
\end{align}
To standardize formulated problem using cost function \eqref{eq22} which is a min-max, auxiliary variables, $\tau(f)$, are defined respecting constraint in
\begin{align}
\tau(f) \in \mathbb{R}^+, \forall e \in E: \tau(f) \geq \dfrac{V(f).\alpha_{fe}}{B^v(e). \beta_{fe}} \label{eq23} \end{align}

Finally, this problem form would be as
\begin{align}
\min \sum_{\forall f \in F}\tau(f) \notag &\\
\text{s.t.} \tau(f) &\geq \dfrac{V(f).\alpha_{fe}}{B^v(e). \beta_{fe}} \label{eq24}
\end{align}

\subsection{Follower Problem Complexity Analysis}
Hessian matrix of the space, which is defined by \eqref{eq24} constraint, is not positive definite and so would not be convexly expressible. In a simplified view of the problem, $\beta_{fe}$ variables could be determined based on that a specific agent selects a specific link to use or not. In this case, optimum shares of the bandwidth are determined easily with a rule of thumb comparing distinct agents' priorities to use network resources. Thus, the simplified problem would be reduced to a selction-based multi-commodity flow problem, with binary variables which itself is reduced to a bin-packing problem, which is proved to be NP-hard.
\subsection{Follower Problem Decomposition}
 A heuristic based on two-phase decomposition is suggested to solve the follower problem. The first step problem ,follower-leader, specifies approximated $\beta_{fe}$s using trivial $\tilde{\alpha}_{fe}$ parameters, whereas the second phase problem, follower-follower, determines approximate $\alpha_{fe}$ parameters using the first phase outputs. Two assumptions are appended to determine trivial  $\tilde{\alpha}_{fe}$ parameters, first that the paths in the network have no or a limited number of joint links, plus that an agent behaves like as there are no other migrating functions present in the network. Consider that single path capacity is determined using its weakest link $B^v(e)$, denoted by $\kappa(p)$. Although a set like $P(n, m)$ shows all the paths in the network between the source $m$ and the destination $n$.
\begin{align}
\widetilde{\alpha}_{fe} &= \dfrac{\sum_{\forall p \in P(S(f), D(f))}^{}[\digamma(p).\Omega(e, p)]}{\sum_{\forall p \in P(S(f), D(f))}^{}[\digamma(p)]} \label{eq25}
\end{align}
\subsection{Follower-Leader Problem}
Based on estimated $\alpha_{fe}$ parameters represented in \eqref{eq25}, the follower-leader problem is formed as expressed in \eqref{eq26}, wherein $\zeta_f = \dfrac{1}{\tau(f)}$ is used to get rid of the quadratic $\tau_f . B^v(e). \beta_{fe} \geq V(f).\widetilde{\alpha}_{fe}$ constraint. This problem aimed to maximize a concave utility function and is easily solvable.
\begin{align}
\max \sum_{\forall f \in F}^{}\big[\dfrac{\zeta_f}{c_u(f) \times R(f)}\big] & \notag \\
\text{s.t.} \ \ \zeta_f &\in \mathbb{R}, \zeta_f > 0, \ \ \forall f \in F, \notag \\
\zeta_f &\leq \dfrac{B^v(e) \beta_{fe}}{V(f)\widetilde{\alpha}_{fe}}, \ \ \forall f \in F, e \in E. \label{eq26}
\end{align}
\subsection{Follower-Follower Problem}
After the first phase, every agents' share of network resources is determined, and so the last phase output, i.e. $\alpha_{fe}$s, are independent of other agents actions, so this last step could be solved distributively regardless of different functions comparison parameters such as $V(f)$ or $C(f)$. This problem is written for an specific agent in \eqref{eq27}, where $\widetilde{\beta}_{fe}$ are the follower-leader problem outputs.
\begin{align}
&\min \tau_e & \notag \\
&\text{s.t.} \ \ \tau_e &\in \mathbb{R}, \tau_e > 0, \forall e \in E \notag \\
&&\tau_e &\geq \dfrac{\alpha_{fe}}{B^v(e) \widetilde{\beta}_{fe}}, \forall e \in E  \label{eq27}
\end{align}
\section{Simulation Results}
\begin{figure}[t]
	\centering
	\includegraphics[width=1\textwidth, scale=0.3, width=75mm]{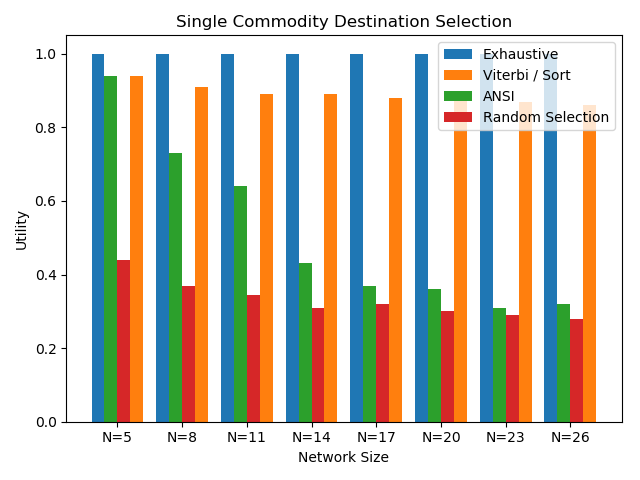}
	\caption
	{
		Various algorithms utility to find single VNF destination in the network -50 percent of the nodes are idle-
	}
	\label{scmf-idle-ttl5}
\end{figure}
In figure \ref{scmf-idle-ttl5}, simulation results for finding a single failed VNF destination in a structured trapezoid network have been shown. In this simulation, idle resources on all of the links are equivalent, and the chance for a server to be idle is 0.5. The right blue column belongs to an exhaustive search. The orange column shows the Viterbi algorithm result, while the green column shows an intuitive handshaking algorithm results, proposed by the authors and named ANSI -not presented in this paper-. The red column shows the utility of random destination selection. Evidently, the Viterbi algorithms performs close to the exhaustive search, and much much better than trivial approach.
\begin{figure}[t]
	\centering
	\includegraphics[width=1\textwidth, scale=0.3, width=75mm]{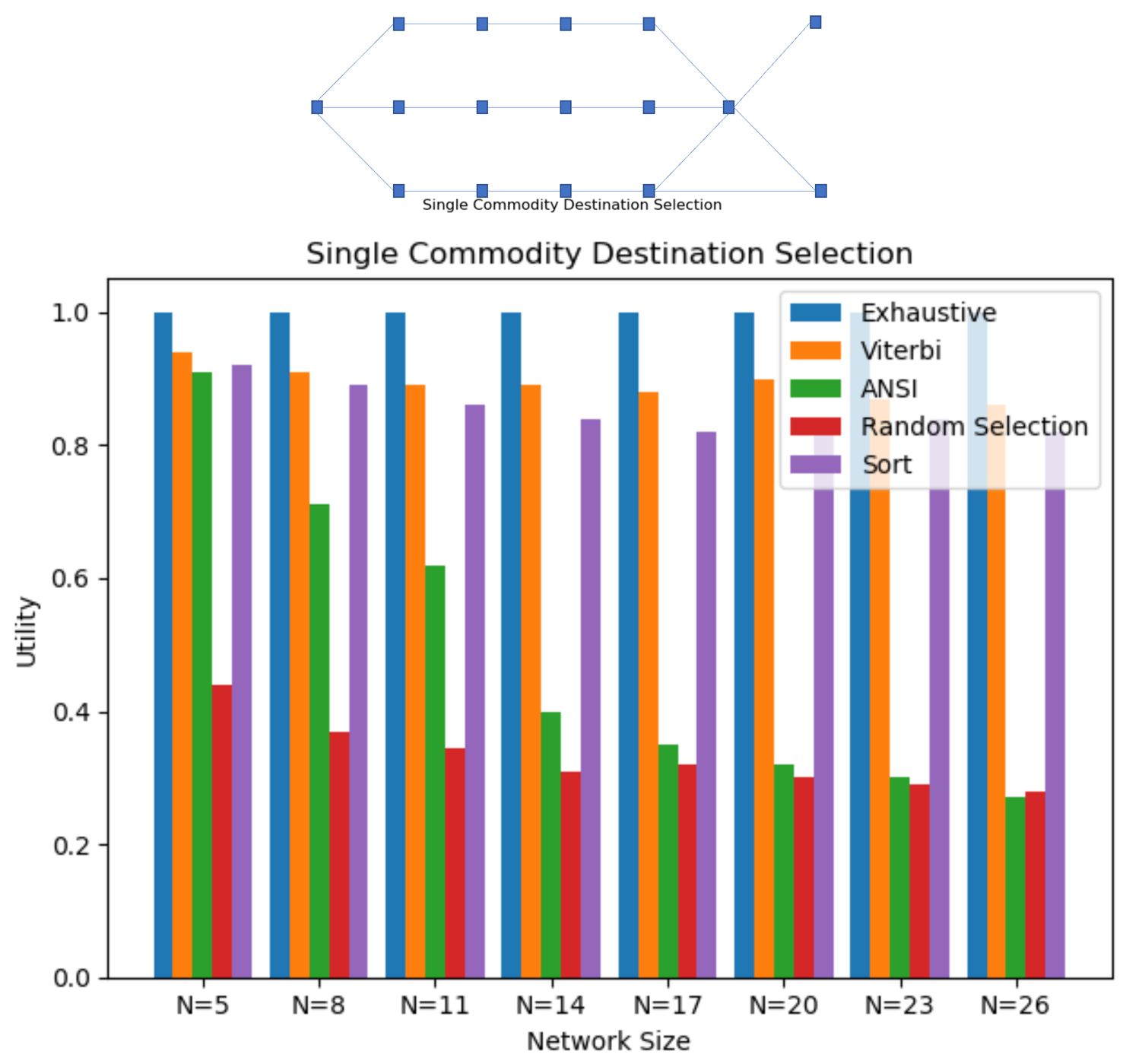}
	\caption
	{
		Various algorithms utility to find single VNF destination in the network -different network nodes load- network structure depicted at the top}
	\label{scmf_load}
\end{figure}
Figure \ref{scmf_load} related run used a slightly different structure than trapezoid -depicted on the top-. This structural change applied to segregate the network utility selection from the idle-avoiding selection in the network. The processing loads on the nodes are uniformly distributed inside $[\dfrac{C_r}{2}, C_r]$ range. Once again, the Viterbi algorithm outperforms other heuristics and remains in a plausible margin of the optimum solution when the network size (the number of the network nodes) increases. -The purple column shows simple candidates sorting based on utility and selection in this figure, obviously, for a single failed VNF this algorithm response qualitys is comparable to the Viterbi-
\begin{figure}[t]
	\centering
	\includegraphics[width=1\textwidth, width=75mm, scale=0.3]{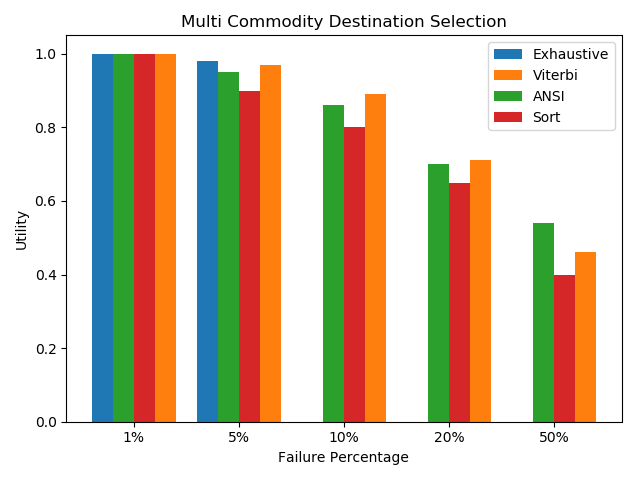}
	\caption
	{
		Various algorithms utilities to find multiple VNF destinations in the network -10$\times$10 network structure-
	}
	\label{mcmf}
\end{figure}
Figure \ref{mcmf} shows the simulation results for finding multiple VNFs destination nodes. The Viterbi outperforms sort and ANSI algorithms in the cases of failure scarcity (failure percentage $\leq 20$), but when failure density increases, the performance of the Viterbi becomes ineligible. Because in this case, hidden states of two successive observations become dependent on actions of more than one function in the network, and the network loses its pseudo-Markov state.

\begin{figure}[t]
	\centering
	\includegraphics[width=1\textwidth, width=75mm, scale=0.3]{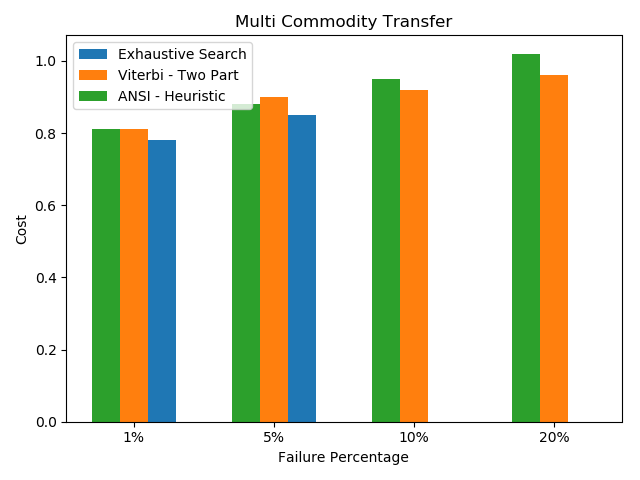}
	\caption
	{
		Different back to back algorithms cost comparison 
	}
	\label{totalsim}
\end{figure}
Figure \ref{totalsim} shows the main problem simulation results, this simulation is done on a 10$\times$10 network, with random links (with probability 0.5 between each pair of nodes). Blue column shows the exhaustive search response, whereas the orange column depicts the result of executing three-phased approaches explained in sections \ref{leader} \& \ref{follower}. Two other back-to-back algorithms for leader and follower problems used to obtain the results shown by the green vector. These algorithms include ANSI as the leader and another heuristic which visualize E2E disjoint paths for different agents as a one-shot follower -Proposed by the authors but not presented here-.
\begin{figure}[t]
	\centering
	\includegraphics[width=1\textwidth, width=75mm, scale = 0.3]{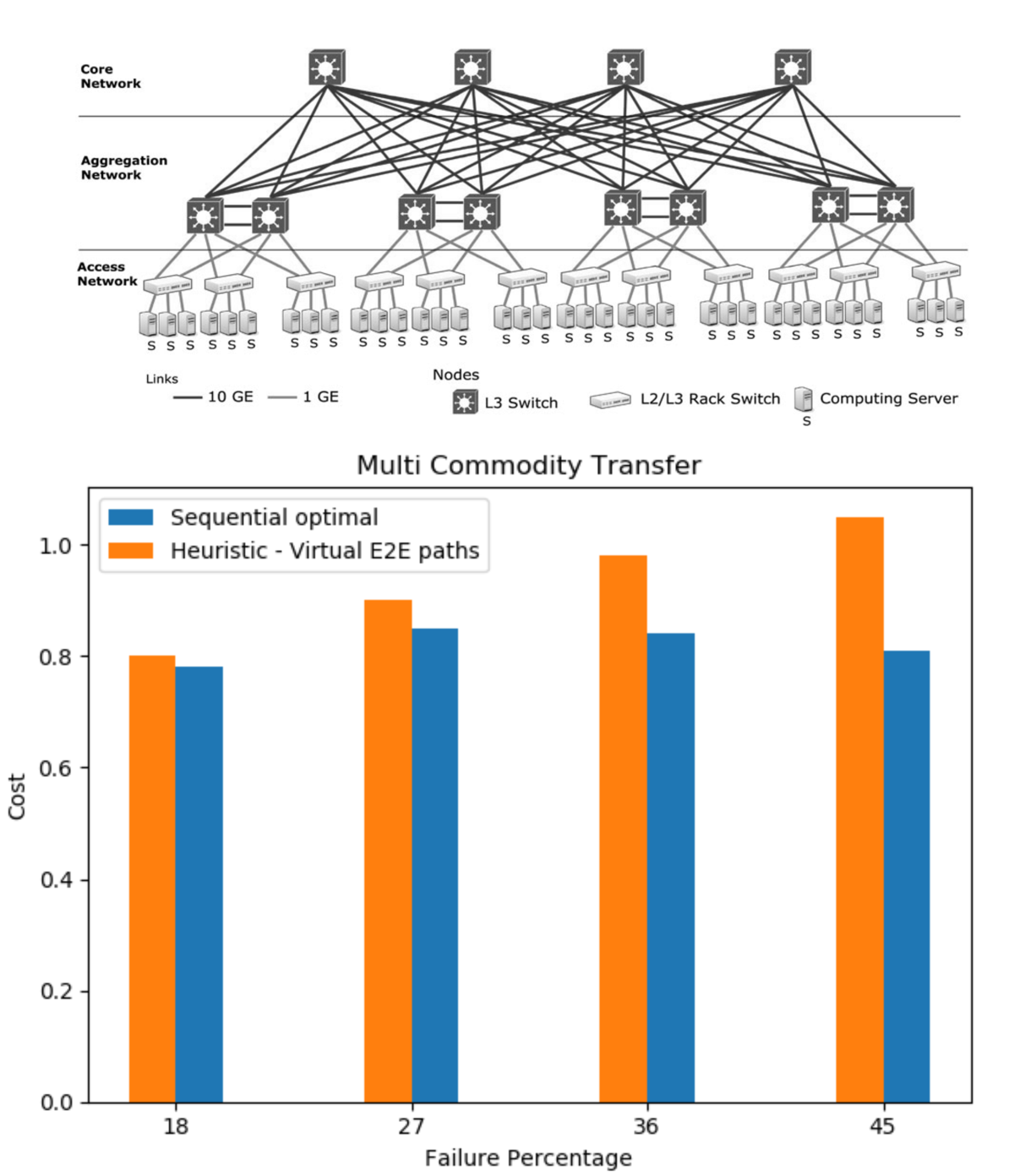}
	\caption
	{
		Cost of using suggested algorithms versus two other back to back heuristic in
		Three-tier network structure -the network structure from \cite{6883798}-
	}
	\label{transfer-tt}
\end{figure}
Figure \ref{transfer-tt} shows the results of solving the primary problem with two different approaches on a three-tier structured network (as explained in \cite{6883798}) with variable size, from 18 to 45 nodes. For this specific structure, the proposed three-phase heuristic (i.e., Viterbi plus follower-leader plus follower-follower)
 outperforms other routine based on sequential executions of the different algorithms explained above. However, in the general case, the selection of the best heuristic depends on the data center network structure. Such severe dependability of the response quality to the network structure is the most important subject that should be noted in future research in this area.

\section{Conclusion}
In this paper, a new look to service functionality protection 
presented. The functions classified into additive and essential types. The impeccable operation of essential functions is guaranteed by using multiple simultaneously active-backup instances. However, additive functions impeccable process is not considered a vital factor in customer quality of experience, and such functions only need to recover in an acceptable limited time after failing to preserve service chain functionality. VNF consistent and integrated state migration to an adequate host in the network after failure is the key to recover VNF functionality. 
in this model. Failed VNF flows forwarding to the central controller during mitigation time and imposed congestion on the settled network traffic encountered as the main migration cost determinant factors in the network. 
Therefore, A system model proposed to investigate cost-effective multiple VNFs state migration in limited time in a generally defined infrastructure network. The resulting optimization problem turned to be MINCP and sub-optimally solved in three phases, each using a heuristic to acquire a distinctive category of problem desirable outputs. The leader 
 is determining VNFs respective destinations 
 using Viterbi algorithm. Then the follower-leader divides network resources between different functions to minimize migration costs and separate different functions migration traffic 
 Finally, the follower-follower problem specifies how a single function state load should be divided among multiple paths to migrate in minimum possible time given the output of two previous phases. 
 Ultimately, simulation results show that the three-phase heuristic responses are close enough to the primitive problem optimum solution.

  \bibliographystyle{IEEEtran}
\bibliography{references}

\end{document}